\documentclass[twocolumn,showpacs,showkeys,preprintnumbers]{revtex4}
\usepackage{graphicx}
\usepackage{subfigure}
\usepackage{lscape}
\usepackage{dcolumn}
\usepackage{bm}
\usepackage{color}

\begin{document}

\title{Diffuse flux of PeV neutrinos from centrifugally accelerated protons in active galactic nuclei}
\author{Rajat K. Dey}
\email{rkdey2007phy@rediffmail.com}
\author{Animesh Basak}
\author{Sabyasachi Ray}
\affiliation{Department of Physics, University of North Bengal, Siliguri, WB 734 013 India}

\begin{abstract}
Evidence for high-energy astrophysical PeV neutrinos has been found in the IceCube experiment from an analysis with a 7.5--year (2010 - 2017) data. Active galactic nuclei (AGN) are among the most prominent objects in the universe, and are widely speculated to be emitters of ultra-high-energy (UHE) cosmic rays with proton domination. Based on the standard two-step LLCD mechanism of particle acceleration, a transformation of energy occurs from AGN's central super-massive black hole (SMBH) rotation to high-energy protons. Protons can be accelerated up to $\sim 0.1$ EeV energies and above, and might generate PeV neutrinos in the energy range $1$--$10$~ PeV through plausible hadronic interactions. The theoretically estimated revised extragalactic diffuse muon neutrino flux employing the {\lq{luminosity-dependent density evolution (LDDE)\rq}} model for the AGN luminosity function (LF) is found consistent with the IceCube level if only a fraction, $6.56\%$ of the total bolometric luminosity (BL) of AGN is being realizable to power the PeV neutrinos. In the $\Lambda$~CDM cosmological framework with the LDDE modeled LF and photon index distribution, about $5.18\%$ of the total BL is enough to power the IceCube neutrinos.
\end{abstract}

\pacs{98.54.Cm, 98.70.Sa, 95.85.Ry} 
\keywords{Active galactic nuclei, Cosmic rays, Neutrinos}
\maketitle

\section{Introduction}
The observation of the ultra-high-energy (UHE) astrophysical neutrinos by the IceCube experiment [1] boosted our theoretical viewpoint to understand their possible origins. The neutrinos can travel largely unhindered through the matter and radiation from their sources. This feature makes them a unique probe that can lead an observer back to their sources. The observed intensity of extragalactic gamma rays in the TeV range, is expected to constrain the UHE neutrino background [2]. However, the absorption en route of these gamma rays in the extragalactic background light sometimes might overturn the direct neutrino-gamma ray connection. 

The $2010\textendash 2017$ data set from the IceCube detector altogether contributed a total of five neutrino events nearly in the $1\textendash 10$ PeV range [3-5]. These PeV energy neutrinos have triggered a serious attention among researchers especially with regard to the exploration of their astrophysical origins. Usually, the hadronic interaction processes like $\rm{p}\gamma$ and $\rm{pp}$ are considered to be in action in different astrophysical sites for the production of these UHE neutrinos [2,6-8]. The active galactic nuclei (AGN) with appropriate bolometric luminosities (BLs) are considered in the present work as probable sources of extragalactic PeV neutrinos. The accretion disk region is likely to be the production site of these neutrinos.    

In the framework of an alternative particle acceleration mechanism, protons and also electrons are boosted to UHEs if energy dissipating  processes could be ineffective during their acceleration phase in the magnetosphere of AGN [9]. These UHE protons with energies $\sim 0.1$~EeV might generate neutrinos through the pion photo-production interactions in the accretion disk zone [10]. Electromagnetic radiation in the UV and soft X-ray band is very abundant ($\rm p_{rad} \gg \rm p_{gas}$, the radiation and gas pressures) in the region of the accretion disk around AGN. The sufficiently energetic protons follow dominant photo-pion production leading to the $\Delta$ resonance in the dense radiative zone. The $\Delta^{+}$ resonance state immediately decays via the following channels:   

\begin{equation}
p + \gamma \rightarrow \Delta^{+} \rightarrow \left\{\begin{array}{ll}
p + \pi^{\rm o} \rightarrow p + 2\gamma \\
n\pi^{+} \rightarrow n + e^{+} + \nu_{\rm e} + \nu_{\mu} + \bar{\nu_{\mu}}
\end{array}
\right. 
\end{equation}

If the above interactions in eq. (1) contribute the PeV neutrino events, there should be a supply of UHE protons with energies all the way close to the range $\sim 0.1$~EeV and above. Such protons (also electrons) could indeed be driven successfully by the AGN through the Landau damping of centrifugally driven Langmuir waves [9]. The pumping of rotational energy in the magnetosphere of AGN into the electric field in the vicinity of the light cylinder surface efficiently supplies the energy for growing Langmuir waves in the bulk electron-proton plasma. The excited Langmuir waves then damp on a faster local proton/electron beam, accelerating them to higher energies [9,11]. The acceleration mechanism is called the Langmuir-Landau-Centrifugal-Drive (LLCD), has been applied to accelerate protons and/or electrons in plasmas surrounding the compact objects (AGN and pulsars) [9,12]. We will not look for the estimation of the predicted gamma ray flux produced simultaneously with neutrinos from $\pi^{0}$-decays in the above process because of their absorption en route from the AGN.

Recalling various acceleration models from earlier works that were exploited to accelerate protons to UHEs in AGN, and ultimately led to power neutrinos observed by the IceCube experiment. In the widely accepted standard acceleration scenario, protons/cosmic rays are believed to be accelerated by internal and re-collimation shocks in the relativistic jets of AGN pointing to us [2,13-14]. With respect to that standard acceleration mechanism, the recently proposed plausible LLCD acceleration mechanism [9] acts as an alternative particle acceleration mechanism in AGN. A little application of an another acceleration model is found in the literature where protons are accelerated by shocks in the cores of AGN ({\it e.g.} Seyfert galaxies) [15]. Recently in [16], authors introduce the AGN corona acceleration model where protons are accelerated stochastically in the coronal plasma by plasma turbulence or magnetic reconnections.

The present work exploits the LLCD mechanism that pumps efficiently the rotational energy of an AGN into the particles' energy. Our main objective in this work, is to investigate primarily the aftermath of the acceleration era of relativistic protons on the soft photons available in the accretion region of rotating AGN with a view to predicting the observed diffuse flux of PeV neutrinos. One should note that the production of PeV neutrinos in the accretion-disk region by centrifugally accelerated protons via LLCD was discussed previously in [17]. The cosmological framework however the former work employed in their calculation for diffuse muon neutrino flux was outdated. First, their work completely overlooked the foremost AGN cosmological evolution, and the inclusion of the cosmological constant or dark energy in the framework of Lambda CDM cosmology. Second, the calculation ignored the effect of the spectral photon index ($\Gamma$) distribution in the number density function of AGN; $f(z,L_{b},{\Gamma})$. With a notable differences from the earlier work, this work has presented a complete calculation of the diffuse PeV muon neutrino flux based on the latest cosmological framework.

We organize the paper as follows: The salient features of the LLCD mechanism for generating the UHE protons is revisited in the next section for consistency. In the third section we estimate the diffuse UHE neutrino flux from the photohadronic interaction channel employing the {\lq{luminosity-dependent density evolution\rq}} model for the AGN luminosity function (LF) with AGN cosmological evolution, while in the last section, our conclusions are summarized. 

\section{LLCD mechanism}

It is believed that the AGN are indeed capable of emitting TeV-PeV energy neutrinos, then the energetics calculation of projectiles is obvious and has already been presented by numerous publications [9,11-13,15, 18-23] and references therein. We will therefore only provide a very brief description of the LLCD mechanism in a standard system like AGN that accelerates protons to energies close to 0.1 EeV and above [9], and finally leads to UHE neutrinos in the PeV range through the photohadronic interactions. It is noteworthy to mention that the recent AGN corona model based on the disk-corona scenario of AGN can explain only the IceCube detected neutrinos in the medium-energy range ($\sim 10\textendash 100$~TeV) [16].

In the first half of the LLCD, the centrifugal acceleration drives the electrostatic Langmuir waves consuming the central SMBH's rotational energy via a parametric two-stream instability with a growth rate [9,11],    

\begin{equation}
\Gamma_{GR}=\frac{\sqrt{3}}{2}{(\frac{\omega_{1}{\omega_{2}}^2}{2})}^{\frac{1}{3}}J_{\mu}(b)^{\frac{2}{3}},
\end{equation}

where $\rm{J}_{\mu}$ is the Bessel's function, and $\rm{b}={(\frac{\omega_{1}}{\Omega})}$, and $\omega_{1,2}= \sqrt{4\pi{\rm{e}^2}\rm{n}_{1,2}/\rm{m_{1,2}}{{\gamma_{1,2}}^3}}$ is the relativistic plasma frequency [9]. ${\rm{n}_{1,2}}$ and ${\gamma_{1,2}}$ are the number density and the Lorentz factor connected with the two species i.e. electrons ($\rm e^{-}$) and protons ($\rm p$). Also, $\rm m_{1,2}$ account their rest masses and $\Omega$ be the angular rate of rotation of AGN's central SMBH. The magnitude of the time dependent centrifugal force that compels the electrostatic waves in the plasma is different for $\rm e^{-}$ and $\rm p$, implying $\gamma_{1}\neq \gamma_{2}$. 

It is obvious that inside the light cylinder zone of an AGN with intense magnetic field ($B$) the $\rm e\textendash p$ plasma co-rotates. The plasma number density in this region is well approximated by the Goldreich-Julian density, $\rm{n}_{GJ}=\rm{\frac{\Omega{B}}{{2\pi{ec}}}}$. An arbitrarily chosen allowed set of  parameters is taken; the BL of AGN, $\rm L_{b}\sim 10^{43}$ erg~s$^{-1}$, $\gamma_{1}\sim 1.6\times 10^{6}, \gamma_{2}\sim 10^{2}$ for calculating the Langmuir instability time-scale, $\frac{1}{\Gamma_{GR}}$. The instability time-scale is found smaller than the kinematic time-scale, $\sim \frac{1}{\Omega}$. Such a condition favors the pumping of rotational energy into Langmuir waves very efficiently in the space.

These waves will impart huge energy to protons by the final step process of LLCD i.e. the Landau damping boosted by a feasible Langmuir collapse. On the contrary, if the index $\mu$ in $\rm J_{\mu}(b)^{\frac{2}{3}}$ is of the order of $\rm b$ or $\rm \frac{2ck~sin\phi_{-}}{\Omega}$, the Langmuir modes have phase velocities for waves exceeding the speed of light corresponding to phase difference, $\phi_{-}>\frac{\pi}{6}$ [9,24]. In this unstable Langmuir modes there are no protons available in the magnetosphere with such speeds, and hence the Langmuir waves will not execute Landau damp anymore.

In [25], authors found in their calculation that in the region with $\rm r<R_{lc}$ ($R_{lc}=\frac{c}{\Omega}$ being the light cylinder radius), the Langmuir waves do not collapse while leaving the magnetospheric region. But, outside the magnetosphere i.e. $\rm r>R_{lc}$, the effects of black hole's rotation/magnetic field upon the plasma kinematics will die out. Instead, the accretion processes solely maintain the particle density in the region. Assuming a spherically symmetric accretion, the estimated proton number density in the region following [9], is

\begin{equation}
n = \frac{L_{b}}{4\eta_{c}\pi{m_{p}}c^2vR_{lc}^2}\approx 6.3\times 10^{5}{(\frac{L_{b}}{10^{42}~erg~s^{-1}})}~cm^{-3}~.
\end{equation}  

We have used $\gamma_2\approx 100$ and $\eta_{c}=0.1$ (only $10 \%$ of the rest mass energy of accretion matter was assumed to give emission). Furthermore, the high frequency pressure parameter, $\rm P_{f}\propto |E|^{2}$ scales as $\frac{1}{l}$ inside the magnetosphere because the strong magnetic field regulates particles to move along 1-dimensional space [25]. But the thermal pressure varies as, $P_{th}\propto\frac{1}{l^{2}}$, in the region and rises faster than $\rm P_{f}$. This behavior envisages the Langmuir collapse not to develop inside the magnetosphere. Outside the magnetosphere, however the $\rm P_{f}$ varies as $\frac{1}{l^{3}}$ and takes over $\rm P_{th}$, ensuring the Langmuir collapse to  occur. 

In [25-26], using $\rm P_{f}$ and $\rm P_{th}$, and approximate solutions of hydrodynamic equations outside the magnetosphere, the dynamic electrostatic field energy was estimated to
\begin{equation}
|W_{E}|^{2} \approx {|E_{0}|^2}{(\frac{t_{0}}{t_{0}-t})^{2}},
\end{equation} 

where $\rm t_{0}$ is the time required for complete collapse of the cavern and $E_{0}$ measures the initial electric field. The caverns are low-density regions, developed from nonlinear instability by the Langmuir turbulence [26]. The sudden rise of the electric field resulting from the dominant pressure component, $\rm P_{f}$, inside the cavern will pull the protons from this space, and efficiently transferring energy from the amplified electrostatic waves to these protons via Langmuir collapse [9].

\subsection{Proton and neutrino energies} 

As a consequence of the final step of the LLCD, the protons are accelerated to desired energy as follows [9],       
\begin{equation}
\epsilon_{p} \approx \frac{ne^{2}}{4\pi^{2}\lambda_{D}^{3}}\Delta{r}^{5}~(eV).
\end{equation}

where $\lambda_{\rm D}$ is the Debye length and $\Delta{\rm r}$ is a narrow  length-scale region in the vicinity of the light cylinder. Taking $\Delta{r}\approx \frac{\rm R_{lc}}{2\gamma_p}$ and $L_{43}\equiv \frac{L_{b}}{(10^{43}~erg~s^{-1})}$, and also $\rm R_{lc,8}\approx 3\times 10^{14}M_8~$cm, the above equation reduces to
\begin{equation}
\epsilon_{p} \approx 1.14\times 10^{17}\times {(\frac{f}{10^{-3}})^3}\times {(\frac{10^2}{\gamma_{p}})^5}\times{(\frac{L_{43}}{M_8})^{5/2}}~(eV), 
\end{equation}

where $\rm f=\frac{\delta{n}}{n_{0}}$ is the initial number density perturbation, and we have chosen $\eta_{c}\approx 0.1$ and $a\approx0.1$ whenever required. 

Here, we assume that each PeV neutrino could receive $\approx 4\%$ on the average of the energy of the proton via the reaction channel in eq. (1) [27].  
\begin{equation}
E_{\nu} \approx 0.04{E_{p}}\approx 2(PeV)~\epsilon_{p,17}[2/(1+z)].
\end{equation} 

Here, $\epsilon_{\rm p}=\epsilon_{p,17}{(10^{17} eV)}$, being the proton energy in the cosmic rest frame and $z$ is the gravitational red-shift of the source. This work includes all AGN in the red shift interval, $0.002 < z\leq 6$ with BLs in a certain range. The very wide and deep field X-ray surveys by {\it{Chandra}}, and {\it{European Space Agency}} directly suggest the existence of a possible large sample of high $\rm z$ AGN with $\rm z> 5$ [28-29]. We, therefore set $\rm z_{max}\approx 6$ in this work.   

We should note here that the protons kinetic (flux) luminosity $\rm{L_{k,p}}$ cannot exceed the $\rm{L_{b}}$ of AGN. It is therefore obvious that, $\rm{L_{k,p}}< \rm{L_{b}}$ and hence the ratio of $\frac{L_{k,p}}{L_{b}}={\eta_{k}}~$ has to be smaller than 1. The interaction probability of $\rm p\gamma$ processes in the accretion region would modify the efficiency of transformation of the $\rm{L_{k,p}}$ to generate UHE neutrinos. Thus, one more parameter ($\rm \xi_{p\gamma}$) has been introduced in order to account the photopion production efficiency of UHE protons with x-ray target photons [16,30]. The parameter $\rm \xi_{p\gamma}$ actually accounts how much fraction of proton's energy carried by pions in the process. For UHE protons with energies $\sim 0.1$~EeV,  the $p\gamma$ process is dominant one over the $pp$ process [31].      

\subsection{Possible energy loss scenario by protons}

It is known that the parametric pumping of Langmuir waves is a highly efficient process. The second step i.e. Langmuir collapse is equally also a fast energy transferring process in systems like AGN. In [25], the initial instability time-scale for Langmuir collapse was measured, and they got a much smaller value compared to the kinematic time-scale ($\frac{2\pi}{\Omega}$). On the other hand, the instability time-scale for Langmuir collapse is comparable to the instability time-scale for Langmuir growth. Very briefly, we will now look upon the possible energy loss mechanisms in the LLCD that may impose significant constraints, if any, during the energy transfer stage to protons [24]. 

The most abundant synchrotron loss mechanism does not affect the continuous energy acquiring mode of protons. In the synchrotron process, the protons could lose their transverse momenta and slide along the magnetic field lines just after their transit to the ground Landau level. In addition, the cooling time-scale of the inverse Compton scattering (ICS) by protons is a continuously increasing (or slow process) function of $\epsilon_{\rm p}$. The ICS mechanism, thus, does not have any role to prevent the LLCD energy transferring. The next possible energy loss process is the curvature radiation, the cooling time-scale of the mechanism in this environment takes much higher values than the overall acceleration time-scale (acceleration or instability timescale is $\sim 10^{-4}$~s [9,11]) and, hence does not interfere notably with the energy transfer process. In the LLCD model the cooling time-scales of $pp$ scattering and Bethe-Heitler (BH) pair production on disk photons do not have noticeable effect to inhibit protons for exceeding photomeson production energies [16]. 

\section{Diffuse neutrino and gamma ray fluxes}

The relativistic protons originating from the close vicinity of the light cylinder of an AGN interact with the x-ray photons present in the accretion disk region described in eq. (1). These processes will lead to the generation of PeV neutrinos and gamma rays. Roughly for every UHE muon neutrino-antineutrino pair generation, there are four twin gamma-ray photons when a significant number of $\rm{p\gamma}$ reactions proceed. Such a variation in the particle numbers arises from the difference between the charge-changing reaction time ($\frac{1}{3}~{\rm{rd}}$) and the charge-unchanging time ($\frac{2}{3}~{\rm{rd}}$) in the possible decay channels of $\Delta^{+}$ state. We shall count this effect to the UHE fluxes of muon neutrinos by the parameter $\rm \chi_{r}$ in the upcoming eq. (20). Then according to the eq. (7), each neutrino will carry an amount of energy, $\rm E_{\nu}\approx\epsilon_{p,17}\frac{4}{(1+z)}$ PeV from an AGN in the appropriate luminosity range. As the charge-unchanging decay mode of $\Delta^{+}$ is more frequent (doubled) than the other channel, each twin gamma-ray photon should therefore carry relatively higher energy compared to a neutrino from the proton via $\pi^{0}$ decay at least in the vicinity of production sites [8]. We do not calculate the diffuse gamma ray flux here because a large fraction of produced PeV gamma rays from the distant AGN in particular, couldn't  reach the Earth due to absorption en route. 

Now, the UHE neutrino flux produced in photohadronic interactions, and subsequent decays, can be estimated theoretically. The UHE neutrino flux ($\rm{{{E_{\nu}}^{2}}\Phi_{\nu}}$) reported by the IceCube experiment in [4,32-33] ranges from $\sim 10^{-9}$ to $\sim 10^{-8}$ GeV cm$^{-2}$s$^{-1}$sr$^{-1}$ corresponding to $E_{\nu}$ in the interval $\sim 0.1$--$10$ PeV. The UHE proton flux coming out of an AGN in terms of cosmic scale factors $R$ in the Robertson-Walker metric (in the Friedmann or FRW cosmological approach) [34] is given by
\begin{equation}
\Phi_{p} = \frac{L_{k,p}R^{2}(t_{1})}{4{\pi}R^{4}(t_{0}){r_{1}}^2}\equiv \frac{{\xi_{p\gamma}}{\eta_{k}}L_{b}R^{2}(t_{1})}{4{\pi}R^{4}(t_{0}){r_{1}}^2},
\end{equation}

where $\rm t_0$ accounts the time when the UHE protons will reach the detector location. The other parameter $\rm t_1$ is the time when these protons left the AGN, and $\rm r_1$ is the corresponding radial distance of the source at that moment. 

Conventionally, the redshift parameter is expressed in terms of the ratio between the scale factors as [34],
\begin{equation}
z = \frac{R(t_0)}{R(t_1)}-1.
\end{equation}

Putting eq. (9) in (8) and after rearrangement of different parameters in the equation, it gives the flux in terms of $z$ as
\begin{equation}
\Phi_{p}(L_{b}) = \frac{{\xi_{p\gamma}}{\eta_{k}}L_{b}}{4{\pi}R^{2}(t_{0}){r_{1}}^2{(1+z)}^2}.
\end{equation}

Now, we introduce a power-law modeled LF in the local universe by $\rm f(z,L_{b})$, that actually estimates the number density of AGN (i.e the comoving density of AGN in some interval of luminosity) [34]. Here, $\rm f(z,L_{b})dL_{b}$ gives the number of AGN per unit volume with luminosities between $\rm L_{b}$ and $\rm L_{b}+dL_{b}$ at redshift $\rm z$ or time $\rm t_1$ [34]. Finally, the UHE flux of protons contributed by all the AGN from distances in the range, $\rm r_{1}:r_{1}+dr_{1}$ with luminosities between $\rm L_{b}:L_{b}+dL_{b}$ is    
\begin{equation}
d\Phi_{p,d} = 4\pi\Phi_{p}(L_{b})R^{2}(t_{1}){r_{1}}^{2}f(t_{1}/z,L_{b})|dt_{1}|\frac{dL_{b}}{L_{*}},
\end{equation}

where $L_{*}$ is called the break luminosity.  

\subsection{Luminosity-dependent density evolution (LDDE) model for $f(z,L_{b})$}

Here, we will describe the cosmological evolution of AGN by taking a  formulation for the AGN LF, given in the eq. (11). The systematic analysis of \emph{Chandra} Deep Field North (CDFN) [35-36] and South (CDFS) [37]  X-ray surveys ($>2$~keV) advocated that the X-ray luminosity function (XLF) of AGN could be well represented with the luminosity-dependent density evolution (LDDE) model [38-40]. As per the LDDE model, the well represented XLF of AGN at a given $z$ is defined using a double power-law LF multiplied a luminosity-dependent evolution term as [29,41]
\begin{equation}
f(z,L_{b}) = \frac{A_{*}}{[(\frac{L_{b}}{L_{*}})^{\gamma_{1}}+(\frac{L_{b}}{L_{*}})^{\gamma_{2}}]}{e(z,L_{b})},
\end{equation} 

The slopes below and above the break luminosity $L_{*}$ are denoted by $\gamma_1$ and $\gamma_2$ in the above. The evolution function is given by 
\begin{equation}
e(z,L_{b})=\left\{\begin{array}{l}
(1+z)^{p_1};\ [z\leq z_{c1}(L_{b})] \\
(1+z_{c1})^{p_1}{(\frac{1+z}{1+z_{c1}})}^{p_2};\ [z_{c1}(L_{b})<z\leq z_{c2}]\\
(1+z_{c1})^{p_1} {(\frac{1+z_{c2}}{1+z_{c1}})}^{p_2}{(\frac{1+z}{1+z_{c2}})}^{p_3};\ [z>z_{c2}]
\end{array}
\right. 
\end{equation}

In the above, $p_1$ to $p_2$ denote the evolution index range corresponding to the cut-off redshift $z_{c1}$, while $z_{c2}$ follows the change from $p_2$ to $p_3$. These redshift cut-offs actually correspond to the redshift where the luminosity evolution changes signs. The luminosity dependence of the index $p_1$ is expressed in [42] as
\begin{equation}
p_{1}(L_{b}) = p_{1}^{*}+\beta_{1}(log{L_{b}}-log{L_{p}}).
\end{equation} 

The cut-off redshifts in terms of luminosity limits are $z_{c1}(L_{b})$ and $z_{c2}(L_{b})$ below and above the luminosity thresholds, $L_{a1}$ and $L_{a2}$, are given by: 
\begin{equation}
z_{c1}(L_{b})=\left\{\begin{array}{ll}
z_{c1}^{*}(\frac{L_{b}}{L_{a1}})^{\alpha_{1}};\      [L_{b}\leq L_{a1}] \\
z_{c1}^{*};\      [L_{b} > L_{a1}]
\end{array}
\right. 
\end{equation}
\begin{equation}
z_{c2}(L_{b})=\left\{\begin{array}{ll}
z_{c2}^{*}(\frac{L_{b}}{L_{a2}})^{\alpha_{2}};\      [L_{b}\leq L_{a2}] \\
z_{c2}^{*};\      [L_{b} > L_{a2}]
\end{array}
\right.
\end{equation}

A good number of best-fit free parameters and their values over the  redshift range $0.002$--$6$ adopting the LDDE model to the XLF of hard X-ray CDF survey are listed in Table 1 [29,43]. The luminosity dependent $p_{1}$ parameter has been taken from a linear fit in the range $\rm log{42}$--$log{44.5}$.

\begin{table*}
	\begin{center}
		\begin{tabular}
			{|l|r|} \hline
			
			{\rm{Parameters}}& {\rm{Values}}\\ \hline
			
			$A_{*}$& ${3.20\pm{0.08}}^{c}$\\ \hline 
			$\gamma_{1}$& $0.96\pm{0.04}$\\ \hline 
			$\gamma_{2}$& $2.71\pm{0.09}$\\ \hline
			$p_{1}$& $3.71$\\ \hline
			$p_{2}$& $-1.5$\\ \hline
			$p_{3}$& $-6.2$\\ \hline 
			$p_{1}^{*}$& $4.78\pm{0.06}$\\ \hline
			$z_{c1}^{*}$& $1.86\pm{0.07}$\\ \hline
			$z_{c2}^{*}$& $3.0$\\ \hline
			$\alpha_{1}$& $0.29\pm{0.02}$\\ \hline
			$\alpha_{2}$& $-0.1$\\ \hline
			$\beta_{1}$& $0.84\pm{0.18}$\\ \hline
			$logL_{*}$& $43.97\pm{0.06}$\\ \hline
			$logL_{a1}$& $44.61\pm{0.07}$\\ \hline
			$logL_{a2}$& $44.0$\\ \hline
			$logL_{p}$& $44.0$\\ \hline    							
		\end{tabular} 
		\footnotesize{ \item[$^{c}$] In $10^{-6}$ ~h$^3_{67.8}$~Mpc$^{-3}$ units; h$_{67.8}\approx 1$.}\\
	\caption {Summary of the best fitted model parameters for XLF in the (2-10) keV x-ray band [29].}
	\end{center}
\end{table*} 

Recalling eq. (11) and using eq. (9) and (10), and $\rm |dt_{1}|=\frac{c}{H_{0}(1+z)^{5/2}}~dz$ in it. 

\begin{equation}
d\Phi_{p,d} = \frac{c{\xi_{p\gamma}}{\eta_{k}}{f(z,L_{b})}{dL_{b}}dz}{(1+z)^{13/2}{H_{0}}{L_{*}}}.
\end{equation}

Now inserting the XLF i.e. eq. (12) in eq. (17), the diffuse flux of UHE protons is given by

\begin{equation}
\Phi_{p,d} = \frac{cA_{*}{\xi_{p\gamma}}{\eta_{k}}}{4{\pi}{H_{0}}{L_{*}}}\int{\frac{L_{b}dL_{b}}{[(\frac{L_{b}}{L_{*}})^{\gamma_{1}}+(\frac{L_{b}}{L_{*}})^{\gamma_{2}}]}\int{\frac{e(z,L_{b})dz}{(1+z)^{13/2}}}}
\end{equation}
in erg~cm$^{-2}$s$^{-1}$sr$^{-1}$.

The luminosity integral is solved numerically corresponding to the luminosity limits of considered AGN in the work. It is already mentioned that the evolution of the LF takes place over the range $\sim 0 \leq \rm z \leq 6$. We have used this upper limit for $\rm z_{max}$ in our calculation. From the eq. (7), we can further ascertain that protons generated via LLCD should gain energies at least in the range $\epsilon_{p,17}\approx 0.25$--$17.5~$, for obtaining IceCube neutrinos in the energy range $1.004$--$10~$ PeV. These protons require a luminosity in the range: $L_{min}$--$L_{max}\approx (0.545$--$2.98)\times 10^{43}$ erg~s$^{-1}$. These limits for $L_{b}$ has yielded: \[\int^{L_{max}}_{L{min}}\frac{{L_{b}}{dL_{b}}}{[(\frac{L_{b}}{L_{*}})^{\gamma_{1}}+(\frac{L_{b}}{L_{*}})^{\gamma_{2}}]}\approx 2\times 10^{87}\]~~in~(erg~s$^{-1}$)$^{2}$.

There are three parts in the redshift integral corresponding to three redshift ranges: $0.002$--$z_{c1}(L_{b})$; $z_{c1}(L_{b})$--$z_{c2}(L_{b})$; $z_{c2}(L_{b})$--$6$, with $z_{c1}(L_{b}) \approx 0.534$; $z_{c2}(L_{b})\approx 3.61$. We take all the useful parameters from the Table 1, and also use eq. (15) to (16) with $L_{b}\approx 10^{42}$ to calculate $z_{c1}(L_{b})$ and $L_{b}\approx 10^{43}$ for $z_{c2}(L_{b})$ (also followed $f$ (in ~Mpc$^{-3}$) versus $z$ curves, in [29]). Hence,\\

$I=\int{\frac{e(z,L_{b})dz}{(1+z)^{13/2}}}=I_{1}+I_{2}+I_{3},$ \\where,

$I_{1}=\int^{0.534}_{0.002}{\frac{(1+z)^{p_{1}}dz}{(1+z)^{13/2}}}\approx 0.294;$\\

$I_{2}=\int^{3.61}_{0.534}\frac{(1+z_{c1})^{p_{1}} {(\frac{1+z}{1+z_{c1}})}^{p_{2}}dz}{(1+z)^{13/2}}\approx 0.0663;$\\

$I_{3}=\int^{6}_{3.61}\frac{(1+z_{c1})^{p_1} {(\frac{1+z_{c2}}{1+z_{c1}})}^{p_2}{(\frac{1+z}{1+z_{c2}})}^{p_3}dz}{(1+z)^{13/2}}\approx 1.32\times 10^{-5}.$\\

The derived flux of UHE protons with above values of integrals, $I_1$, $I_2$ and $I_3$ is
\begin{equation}
\Phi_{p,d} \approx 7.2^{+1.81}_{-1.41} \times 10^{86}\times\frac{cA_{*}{\xi_{p\gamma}}{\eta_{k}}}{{4{\pi}H_{0}}{L_{*}}}
\end{equation}

in erg~cm$^{-2}$s$^{-1}$sr$^{-1}$.

Hence, the muon neutrino flux contributed by all the AGN with $z=0.002$--$6$ and $L_{b}= (0.545$--$2.98)\times 10^{43}$~erg~s$^{-1}$ is 
\begin{equation}
\Phi_{\nu_{\mu}/{\gamma}} \approx 4.46^{+1.12}_{-0.87} \times 10^{89}\times\frac{cA_{*}{\xi_{p\gamma}}{\eta_{k}}\chi_{r}\chi_{o}}{4{\pi}{H_{0}}{L_{*}}}
\end{equation}

in GeV~cm$^{-2}$s$^{-1}$sr$^{-1}$.\\

The parameter $\chi_{r}$ takes $\frac{2}{3}$ for the calculation of muon neutrino flux [7]. The final products of neutrino flavors maintain the conversion as; $\nu_{\rm e}:\nu_{\mu}:\nu_{\tau}=1:2:0$ near the production site but suffers a transition (resulting from maximal mixing of $\nu_{\mu}$ and $\nu_{\tau}$ due to neutrino oscillations) into, $\nu_{\rm e}:\nu_{\mu}:\nu_{\tau}\approx 1:1:1$ at the observation point [44-45]. Hence, $\chi_{o} = \frac{1}{2}$ is appropriate for muon neutrinos. The photopion production efficiency in the energy range $\epsilon_{p}\sim 10^{16}$--$10^{18}$ is expected to be $\xi_{p\gamma}\approx 0.20$ [14,30] corresponding to considered luminosity limits of AGN here. We take Hubble's constant, $\rm H_{0}\approx 67.8~$km~s$^{-1}$~Mpc$^{-1}$ [46-47].

Taking data from the Table 1, and using other relevant parameters available in the paper, the PeV muon neutrino flux resulting from the theory with the LDDE model for the XLF, is $3.98^{+0.46}_{-0.38}\times 10^{-8}\eta_{k}$ in GeV~cm$^{-2}$s$^{-1}$sr$^{-1}$.

\subsection{Luminosity-dependent density evolution (LDDE) model with photon index distribution: $f(z,L_{b},\Gamma)$}

Up till now, we have employed the FRW cosmological framework that did not include the cosmological constant ($\Lambda$) yet. It has been studied later that the presence of the $\Lambda$ in the FRW model (also, called Lambda CDM model) has played an important role in the study of the universe. For the cosmology, we take the parameters ($\Omega_{m},~\Omega_{\Lambda})=(0.3,0.7)$ in the calculation of the diffuse flux of muon neutrinos. Starting from eq. (18) and using the standard $\Lambda{CDM}$ cosmological framework, the diffuse muon neutrino flux contributed by all the AGN in the universe at $z=0.002$--$6$ is given by
\begin{equation}
\Phi_{\nu_{\mu}/{\gamma}} = \frac{cA_{*}{\xi_{p\gamma}}{\eta_{k}}\chi_{r}\chi_{o}}{4{\pi}{H_{0}}{L_{*}}}I_{L}I_{z}I_{\Gamma}
\end{equation}
where,~\\

$I_{L}=\int^{L_{max}}_{L_{min}}{\frac{L_{b}dL_{b}}{[(\frac{L_{b}}{L_{*}})^{\gamma_{1}}+(\frac{L_{b}}{L_{*}})^{\gamma_{2}}]}};$\\

$I_{z}=\int^{z_{max}}_{z_{min}}{\frac{e(z,L_{b})dz} {(1+z)^5{[\Omega_{m}(1+z)^{3}+\Omega_{\Lambda}]^{1/2}}}};$\\

$I_{\Gamma}=\int^{\Gamma_{max}}_{\Gamma_{min}}\frac{dN}{d\Gamma}{d\Gamma}.$\\ 

We have used, $\frac{dt_{1}}{dz}=\frac{c}{H_{0}(1+z){[\Omega_{m}{(1+z)}^{3}+{\Omega_{\Lambda}}]}^{1/2}}$ in the above. The first integral has already been evaluated above and its value is $\approx 2\times 10^{87}$~~in~(erg~s$^{-1}$)$^{2}$. Following the eq. (13), the integral $I_{z}$	is a sum of three integrals maintaining three different ranges of redshift limits as before but having different integrands. We have evaluated them numerically and their sum is $I_{z}=I_{1}+I_{2}+I_{3}\approx 0.4555$. In $I_{\Gamma}$, the function $\frac{dN}{d\Gamma}$ describes the photon index distribution and is assumed to be free from $z$. In [29,48], it was modeled as a Gaussian function:

\begin{equation}
\frac{dN}{d\Gamma}=\frac{1}{\sigma_{\Gamma}\sqrt{2\pi}}exp[-\frac{({\Gamma}-{\mu})^2}{2\sigma_{\Gamma}^2}]
\end{equation}

where $\mu$ and $\sigma_{\Gamma}$ have values 1.88 and 0.15, obtained from the best fitting of the spectra of the Swift/BAT sample with the $\Gamma$ range; $\Gamma_{min}=1.40$ and $\Gamma_{max}=2.35$ [29,49]. Using these limits and best fit parameters to the integral of the distribution of photon index, the $I_{\Gamma}$ yields a value $\approx 0.9984$.

Using eq. (21), the derived diffuse PeV muon neutrino flux from all the AGN is    
\begin{equation}
\Phi_{\nu_{\mu}/{\gamma}} \approx 5.03^{+0.58} _{-0.48}\times 10^{-8}{\eta_{k}}
\end{equation}

in GeV~cm$^{-2}$s$^{-1}$sr$^{-1}$.\\

\subsection{Constraints on BL by the IceCube limit on measured neutrino flux}

The IceCube neutrino flux obeys a best-fit power law as, is $\rm J_{\nu}(E_{\nu})={E_{\nu}}^{2}\Phi_{\nu}(E_{\nu})\approx (2.2\pm{0.8})\times 10^{-8}(\frac{E_{\nu}}{0.1~PeV})^{-0.91}~$GeV~cm$^{-2}$~s$^{-1}$~sr$^{-1}$ [50]. We can describe the energy spectra of neutrinos by $\rm J_{\nu}(E_{\nu})dE_{\nu}={E_{\nu}}dN_{\nu}$, where $\rm dN_{\nu}=\Phi_{\nu}(E_{\nu})dE_{\nu}$, being the number of neutrinos in the energy interval, $\rm E_{\nu}$:$\rm E_{\nu}+dE_{\nu}$. The diffuse flux of muon neutrinos observed at Earth in the energy range, $\approx~1$--$10$ PeV could be [50]

\begin{equation}
\Phi_{\nu_{\mu},d}=\int^{10}_{1}{E_{\nu}}^{-1}J_{\nu}(E_{\nu})dE_{\nu}\approx (2.61\pm{0.95})\times 10^{-9}\\
\end{equation}

in GeV cm$^{-2}$s$^{-1}$sr$^{-1}$.\\

The calculated extragalactic muon neutrino flux may lead to constrain on the percentage of conversion of the BL to power the IceCube's measured PeV muon neutrino flux.  For the LDDE modeled XLF, it is found that the model predicted flux could give the IceCube measured PeV neutrino flux per flavor if $\eta_{\rm k}$ takes a value $\approx 6.56\%$. Finally, exploiting the $\Lambda~$CDM cosmology along with the photon index distribution in the LDDE modeled XLF, the factor $\eta_{k}$ becomes $\approx 5.18\%$.

In the accretion disk region, UHE gamma rays proceed via the following channel: $\rm p+\gamma_{soft} \rightarrow \pi^{0} \rightarrow \gamma \gamma \rightarrow e^{+}e^{-}$. The optical depth in the radiation field is least along the axis of the accretion disk, TeV gamma rays (converted from PeV gamma rays due to internal $\gamma\gamma$ annihilation) and a tiny fraction of PeV gamma rays may escape along the disk axis [51].

The fraction of the PeV gamma rays that can escape from the AGN along its accretion disk axis, could not survive against strong absorption caused by the extragalactic background radiations. Their cosmological distances make them impossible to reach at Earth. For gamma rays with energies, $\geq 1~$PeV, the interaction mean free path due to $e^{+}e^{-}$ production on CMB is $\geq 10$ kpc only [52]. In agreement with the above conjecture, no extragalactic PeV gamma rays or gamma ray sources (AGN) are going to be detected in future [53]. 

\section{Conclusions}

In the present work, we have investigated the generation of PeV neutrinos ever detected through a specific photohadronic interaction channel in the vicinity of AGN's SMBH. A brief summary of our  conclusions is the following.

We have highlighted some salient features of the LLCD mechanism in connection with the acceleration of protons in the magnetospheric plasma of AGNs. Acceleration of protons proceeds via two successive steps: {\bf i.} Langmuir waves are generated due to electrostatic field caused by the black hole's rotation and {\bf ii.} a systematic collapse of Langmuir waves on the local beam of relativistic protons leading to their further acceleration in the vicinity of the light cylinder. 

If protons reach to a medium energy, $\sim 0.1$ EeV in the magnetosphere of AGN, they interact with enough target photons (soft X-ray and UV radiation fields) in the accretion disk region of the AGN, producing neutrinos with energies in the range $1$--$10~$PeV.

The estimated diffuse muon neutrino flux using the $\Lambda$-CDM cosmology for the AGN LF is found consistent with one that has been predicted by erstwhile standard FRW cosmology with AGN cosmological evolution. Corresponding to these two scenarios about $6.56\%$ and $5.18\%$ of the total BL of all considered AGN are just enough to interpret the PeV energy neutrino flux observed by IceCube. A single-power law LF without the AGN cosmological evolution, a recent computation based on LLCD found a small fraction $\approx 0.003\%$ of the total BL to power the IceCube's PeV neutrinos.  

In recent past many works assume that the origin of PeV neutrinos might be some classified AGN {\it {viz.}} blazars, BL Lacs, flat spectrum radio quasars (FRSQs) and some other astrophysical objects {\it {viz.}} gamma-ray bursts, Type IIn supernovae (SNRs), pulsars, magnetars etc. The computed diffuse PeV neutrino fluxes $E_{\nu}^2\phi_{\nu}$ from FRSQs and BL Lacs were $ < 7.4\times 10^{-10}$ and $< 9.94\times 10^{-11}$ in GeV cm$^{-2}$s$^{-1}$sr$^{-1}$ [48,54]. Based on the nonlinear diffusive shock acceleration model, contribution from Type IIn SNRs is $\sim 10^{-9} - 10^{-8}$ [55]. Moreover, the recently proposed AGN corona model could possibly reveal only the origin of the medium-energy ($\sim 10 - 100$~TeV) neutrinos observed by the IceCube.      

The extragalactic diffuse PeV gamma rays produced simultaneously with neutrinos are unlikely to be available at Earth because of their strong absorption en route.  

\section*{Acknowledgment}
Authors would like to thank the anonymous Reviewers for their stimulating critique that helped improve the manuscript. RKD acknowledges the financial support from North Bengal University under the Teachers' Research Project Scheme; Ref.No. 1513/R-2020.


\begin{thebibliography}{99}
\bibitem{rd:1} Aartsen M. G. \emph{et al.}, \emph{Phys. Rev. Lett.}, \textbf{111} (2013) 021103. 
\bibitem{rd:2} Cholis I. and Hooper D., \emph{JCAP}, \textbf{06} (2013) 030.
\bibitem{rd:3} Aartsen M. G. \emph{et al.}, \emph{Phys. Rev. Lett.}, \textbf{113}  (2014) 101101. 
\bibitem{rd:4} Aartsen M. G. \emph{et al.}, \emph{Science}, \textbf{342} (2013) 1242856.
\bibitem{rd:5} Niederhausen H., \emph{18th Conference on Elastic and Diffractive Scattering}, Vietnam, (2019) arXiv:1909.12182v2.
\bibitem{rd:6} Roulet E. \emph{et al.}, \emph{JCAP}, \textbf{1301} (2013) 028.
\bibitem{rd:7} Bhadra A. and Dey R. K., \emph{MNRAS}, \textbf{395} (2009) 1371.
\bibitem{rd:8} Dey R. K. \emph{et al.}, \emph{EPL}, \textbf{115} (2016) 69002 .
\bibitem{rd:9} Osmanov Z. \emph{et al.}, \emph{MNRAS}, \textbf{445} (2014) 4155.
\bibitem{rd:10} Shakura N. I. and Sunyaev R. A., \emph{Astron. and Astrophys.}, \textbf{24} (1973) 337.
\bibitem{rd:11} Machabeli G.\emph{et al.},\emph{Phys. Plasmas}, \textbf{12} (2005) 062901.
\bibitem{rd:12} Osmanov Z. \emph{et al.}, \emph{Nat. Sci. Rep.}, \textbf{5} (2015) 14443.
\bibitem{rd:13} Atoyan A. and Dermer C., \emph{Phys. Rev. Lett.}, \textbf{87} (2001) 221102.
\bibitem{rd:14} Mannheim K. \emph{et al.}, \emph{Phys. Rev. D.}, \textbf{63}, (2001) 023003.
\bibitem{rd:15} Stecker F. W., \emph{Phys. Rev. D.}, \textbf{88} (2013) 047301.
\bibitem{rd:16} Murase K. \emph{et al.}, \emph{Phys. Rev. Lett.}, \textbf{125} (2020) 011101.
\bibitem{rd:17} Osmanov Z. \emph{et al.}, \emph{Astropart. Phys.}, \textbf{99} (2018) 30.
\bibitem{rd:18} Alvarez-Muniz J. and Meszaros P., \emph{Phys. Rev. D}, \textbf{70} (2004) 123001.
\bibitem{rd:19} Waxman E. and Bahcall J. N., \emph{Phys. Rev. Lett.}, \textbf{78} (1997) 2292.
\bibitem{rd:20} Dermer C. D., \emph{Astrophys. J.}, \textbf{574} (2002) 65.
\bibitem{rd:21} Murase K. \emph{et al.}, \emph{Phys. Rev. D}, \textbf{79} (2009) 103001.
\bibitem{rd:22} Levinson A. and Waxman E., \emph{Phys. Rev. Lett.}, \textbf{87} (2001) 171101.
\bibitem{rd:23} Murase K. \emph{et al.}, \emph{Phys. Rev. D}, \textbf{84} (2011) 043003.
\bibitem{rd:24} Mahajan S. \emph{et al.}, \emph{Nat. Sci. Rep.}, \textbf{3} (2013) 1262.
\bibitem{rd:25} Artsimovich L. A. and Sagdeev R. Z., \emph{Plasma Physics for Physicists} (1979) (Atomizdat: Moscow).
\bibitem{rd:26} Zakharov V. E., \emph{Sov. J. Exp. Theor. Phys.}, \textbf{35} (1972) 908.
\bibitem{rd:27} Murase K. \emph{et al.}, \emph{Phys. Rev. D}, \textbf{88} (2013) 121301.
\bibitem{rd:28} Barger A. J. \emph{et al.}, \emph{Astrophys. J.} \textbf{124} (2002) 1839.
\bibitem{rd:29} Ueda Y. \emph{et al.}, \emph{Astrophys. J.}, \textbf{786} (2014) 104.
\bibitem{rd:30} K. Murase \emph{et al.}, \emph{Phys. Rev. D}, \textbf{78} (2008) 023005.
\bibitem{rd:31} Zhang B. T. and Li Z., \emph{JCAP}, \textbf{03} (2017) 024.
\bibitem{rd:32} Aartsen M. G. \emph{et al.}, \emph{EPJ C}, \textbf{77}  (2017) 692.
\bibitem{rd:33} Aartsen M. G. \emph{et al.}, \emph{Nature}, \textbf{591} (2021) 220.
\bibitem{rd:34} Weinberg S., \emph{Gravitation and Cosmology: Principles and Applications of the General Theory of Relativity}, \emph{John Wiley and Sons Inc}, (2008).
\bibitem{rd:35} Alexander D. M. \emph{et al.},  \emph{Astron. J.}, \textbf{126} (2003) 539.
\bibitem{rd:36} Trouille L. \emph{et al}, \emph{Astrophys. J. Suppl.}, \textbf{179} (2008) 1.
\bibitem{rd:37} Xue Y. Q. \emph{et al.}, \emph{Astrophys. J. Suppl.}, \textbf{195} (2011) 10.
\bibitem{rd:38} Ueda Y. \emph{et al.}, \emph{Astrophys. J.}, \textbf{598} (2003) 886.
\bibitem{rd:39} Ebrero J. \emph{et al.}, \emph{Astron. and Astrophys.}, \textbf{493} (2009) 55.
\bibitem{rd:40} Yencho B. \emph{et al.}, \emph{Astrophys. J.}, \textbf{698} (2009) 380.
\bibitem{rd:41} Hopkins P. F. \emph{et al.}, \emph{Astrophys. J.}, \textbf{654} (2007) 731.
\bibitem{rd:42} Hasinger G. \emph{et al.}, \emph{Astron. and Astrophys.}, \textbf{441} (2005) 417.
\bibitem{rd:43} Fiore F. \emph{et al.}, \emph{Astron. and Astrophys.}, \textbf{537} (2012) 16.
\bibitem{rd:44} Gonzalez-Garcia M. C., \emph{JHEP}, \textbf{11} (2014) 052.
\bibitem{rd:45} Athar H. \emph{et al.}, \emph{MPLA}, 21 (2006) 1049.
\bibitem{rd:46} Ade P. A. R. \emph{et al.}, \emph{Astron. and Astrophys.}, \textbf{571} (2014) 48.
\bibitem{rd:47} Ade P. A. R. \emph{et al.}, \emph{Astron. and Astrophys.}, \textbf{594} (2016) A13.
\bibitem{rd:48} Ajello M. \emph{et al.}, \emph{Astrophys.J.}, \textbf{751} (2012) 108.
\bibitem{rd:49} Burlon D. \emph{et al.}, \emph{Astrophys. J.} \textbf{728} (2011) 58.
\bibitem{rd:50} Williams D., \emph{IJMP Con. Ser.}, vol. \textbf{46} (2018) 1860048.
\bibitem{rd:51} Zhang L. and Cheng K. S., \emph{Astrophys.}, \textbf{488} (1997) 94.
\bibitem{rd:52} Protheroe R. J. and Biermann P. L., \emph{APh}, \textbf{6} (1996) 45.
\bibitem{rd:53} Ahnen M. L. \emph{et al.}, \emph{Astron. and Astrophys.}, \textbf{595} (2016) A98.
\bibitem{rd:54} Ajello M. \emph{et al.}, \emph{Astrophys. J.}, \textbf{780} (2014) 73.
\bibitem{rd:55} Zirakashvili V. N. and Ptuskin V. S., \emph{Astropart. Phys.}, \textbf{78} (2016) 28.
\end{thebibliography}
\end{document}